\documentclass[preprint,showpacs,preprintnumbers,amsmath,amssymb]{revtex4}
\usepackage{graphicx}
\usepackage{dcolumn}
\usepackage{bm}
\usepackage{epsf}

\renewcommand{\bar}[1]{\overline{#1}}

\renewcommand{\d}{{\mathrm d}}

\usepackage{amssymb}

\renewcommand{\d}{{\mathrm d}}
\renewcommand{\bar}[1]{\overline{#1}}

\usepackage{indentfirst}
\usepackage{psfig,color}
\usepackage{epsfig}
\usepackage{epsf}
\usepackage{graphicx}
\begin{document}

%\begin{flushright}
%\end{flushright}

\title{Effect due to charge symmetry violation on the Paschos-Wolfenstein relation }

\author{Yong Ding}
\affiliation{Department of Physics, Peking University, Beijing
100871, China}
\author{Bo-Qiang Ma}
\email{mabq@phy.pku.edu.cn} \altaffiliation{corresponding author.}
\affiliation{ CCAST (World Laboratory), P.O.~Box 8730, Beijing 100080, China\\
Department of Physics, Peking University, Beijing 100871,
China\footnote{Mailing address.}}

\begin{abstract}
The modification of the Paschos-Wolfenstein relation is
investigated when the charge symmetry violations of valence and
sea quark distributions in the nucleon are taken into account. We
also study qualitatively the impact of charge symmetry violation
(CSV) effect on the extraction of $\sin^{2}\theta_{w}$ from deep
inelastic neutrino- and antineutrino-nuclei scattering within the
light-cone meson-baryon fluctuation model. We find that the effect
of CSV is too small to give a sizable contribution to the NuTeV
result with various choices of mass difference inputs, which is
consistence with the prediction that the strange-antistrange
asymmetry can account for largely the NuTeV deviation in this
model. It is noticeable that the effect of CSV might contribute to
the NuTeV deviation when the larger difference between the
internal momentum scales, $\alpha_{p}$ of the proton and
$\alpha_{n}$ of the neutron, is considered.
\end{abstract}


%\vfill

%\preprint{Submitted to Phys. Rev. D }

%\preprint{To appear in PRD}

%Latex2e

%\vspace{3cm}

\vfill


%\vspace{5cm}

\vfill

%\vspace{3cm}

%\vfill


%\keywords{Suggested keywords}%Use show keys class option if keyword

\vfill %\vspace

%{\centerline{ Accepted by Phys. Rev. D for publication}}
%\vspace

%\vfill

\maketitle

\section{Introduction}

In recent years, the precise determination of weak-mixing angle (or
Weinberg angle) $\sin^{2}\theta_{w}$ has received a lot of
attention. It is well known that the Weinberg angle is one of the
key parameters in the standard model~(SM) of electroweak theory and
can be determined from various experimental methods, such as atomic
parity violation, $W$ and $Z$ masses, elastic and inelastic neutrino
scattering, and so on. In 2002, the NuTeV
Collaboration~\cite{zell02} announced that they measured a new
value:
$\sin^{2}\theta_{w}=0.2277\pm0.0013~(\mbox{stat})\pm0.0009~(\mbox{syst})$,
which is larger than the world accepted value:
$\sin^{2}\theta_{w}=0.2227\pm0.0004$ measured in other electroweak
processes with three standard deviations. NuTeV extracted the value
of $\sin^{2}\theta_{w}$ by measuring the ratio of
 neutral-current to charged-current cross
sections for neutrino and anti-neutrino on the iron targets,
respectively, and then made a full Monte-Carlo simulation of their
experiment. A number of corrections should be considered before
any conclusion may be drawn, because the analysis procedure is
based on the Paschos-Wolfenstein~(P-W) relation~\cite{pash73}
\begin{equation}
R^{-}=\frac{\sigma^{{\nu}N}_{NC}-\sigma^{\overline{\nu}N}_{NC}}{\sigma^{{\nu}N}_{CC}
-\sigma^{\overline{\nu}N}_{CC}}=\frac{1}{2}-\sin^{2}\theta_{w}.
\label{ratio}
\end{equation}
In this equation, $\sigma^{\nu N}_{NC}$~($\sigma^{\bar{\nu}
N}_{NC}$) is the integral of neutral-current inclusive
differential cross section for neutrino~(antineutrino) over $x$
and $y$, and it is similar to $\sigma^{\nu
N}_{CC}$~($\sigma^{\bar{\nu} N}_{CC}$). This relation provides an
independent determination of the Weinberg angle. Three assumptions
should be made for the validity of this relation: isoscalar
target, which means that the number of protons is equal to that of
neutrons for the target; quark-antiquark symmetries for both
strange and charm quark distributions
($s(x)=\bar{s}(x)$,$c(x)=\bar{c}(x)$); and charge
symmetry~($u_{p}(x)=d_{n}(x)$,$d_{p}(x)=u_{n}(x)$ and similarly
for $\bar{u}(x),\bar{d}(x)$), where $x$ represents the momentum
fraction carried by the quark in the nucleon. In fact, these
assumptions are not strictly valid in realistic reactions.
Usually, there is a small deviation from an isoscalar target by an
excess of neutrons over protons, which has been considered by the
NuTeV Collaboration. But the NuTeV Collaboration disregarded not
only the effect due to strange-antistrange asymmetry but also the
CSV effect in their original analysis. Although many sources of
systematic errors and several uncertainties have been considered,
it is still an open question whether the NuTeV deviation could be
accounted for within or beyond SM. Possible sources of the NuTeV
anomaly beyond SM have been discussed in Ref.~\cite{davidson02}.
However, before speculating on the possible new physics, one
should first check carefully the {\it standard} effect and the
theoretical uncertainties coming from complicated aspects of the
quantum chromodynamics (QCD).

As mentioned above, one of the assumptions is the isoscalar target,
i.e., the nucleus should be in an isoscalar state, so that various
strong interaction effects can cancel out in the ratio. However, the
targets used in the neutrino experiments are usually non-isoscalar
nuclei with a significant neutron excess, such as the iron target in
the NuTeV experiment. The corrections of non-isoscalar target to the
NuTeV anomaly were given in
Ref.~\cite{k02,k0412307,kulagin03,ksy02}. Besides that there are
other suggestions~\cite{0204007} from a conservative point of views.

The second assumption is the quark-antiquark symmetry of strange
and charm momentum distributions in the nucleon sea. The validity
of these asymmetries for quark-antiquark has been discussed by a
lot of investigators not only for its connection to the proton
spin problem but also for the probability to explain the anomalous
value of the Weinberg angle. The asymmetry of strange and
antistrange momentum distributions is the most sensible
explanation for the NuTeV anomaly within SM.
%, at least on the surface.
Recently, the contribution caused by the strange-antistrange
asymmetry in perturbative quantum chromodynamics (PQCD) at
three-loop~\cite{pQCD} was predicted, but it is too small to affect
the extraction of the weak-mixing angle. Thus, the reason for the
asymmetric momentum distributions of strangeness in the nucleon sea
should be of nonperturbative origin~\cite{bm96,ST,bw}. In fact,
there have been a series of discussions on the effect due to
strange-antistrange asymmetry on the extraction of
$\sin^{2}\theta_{w}$. Cao and Signal~\cite{cs03} investigated this
asymmetry by using the meson cloud model and found that the result
is fairly small with almost no correction to the value of
$\sin^{2}\theta_{w}$. Also, in the past, the role of asymmetric
strange-antistrange quark momentum distributions was predicted by
using the light-cone meson-baryon fluctuation model~\cite{dm04} and
the effective chiral quark model~\cite{dxm04,dxm05}, respectively.
Noticeably, the similar predictions that the effect of $s$-$\bar{s}$
asymmetry would remove largely the NuTeV anomaly were obtained from
the above two different models. In addition, in Ref.~\cite{dxm05},
the strangeness asymmetries were compared between the prediction by
the effective chiral quark model and the parameterizations of the
NuTeV experimental data. It was found that the prediction of this
model is consistent with the parameterizations of the experimental
data. The same conclusion was given in Ref.~\cite{wakamastu04} with
the chiral quark soliton model by introducing a parameter of the
effective mass difference between strange and nonstrange quarks, an
idea also discussed in Ref.~\cite{li02}. Alternatively, Olness
$et.~al.$~\cite{olness03} performed the first global QCD analysis of
the CCFR and NuTeV dimuon data, adopted a general parameterization
of the nonperturbative $s(x)$ and $\bar{s}(x)$ distribution
functions, and evaluated the contribution to the NuTeV deviation
with uncertainties. Furthermore, the correction coming from QCD to
the shift of $\sin^{2}\theta_{w}$ which relates with the effects of
isospin violation and asymmetry of strangeness content were also
estimated in Ref.~\cite{kretzer04}.

Despite so many theoretical arguments and analyses, e.g.,
Ref.~\cite{alwall} and Refs.~\cite{b95,sbr97,a97}, there is no
direct experimental evidence for the asymmetric momentum
distributions of strange quark and antiquark in the nucleon sea
except the dimuon experiment induced by neutrinos and
antineutrinos~\cite{b95} which is the best method for measuring
the $s$-$\bar{s}$ asymmetric momentum distributions at this stage.
The precision of the dimuon experiment is not high enough to get
the detailed information about this asymmetry. Moreover, the
effect of strange-antistrange asymmetry often mixes with the CSV
effect in experiment. But it is still possible to extract the
effects of asymmetric strangeness distributions and CSV from
different experiments in the future~\cite{gao05}.

The so-called charge symmetry is the invariance of the QCD
Lagrangian with the up ($u$) and down ($d$) quarks interchanging
when both the mass difference of them and the electromagnetical
effects are ignored. This invariance is a more restricted form of
isospin invariance involving a rotation of $180^{\circ}$ about the
``2" axis in isospin space, or more specifically, the isospin
symmetry for the $u \leftrightarrow d$ exchange between the proton
and the neutron. Thus, the light flavor
%quark distributions in the nucleon can be extracted from data by a factor of two, i.e., the
parton distributions in the neutron can be expressed in terms of
those in the proton. It should be emphasized that the charge
symmetry violation effect arises from the mass difference between up
and down quarks and from electromagnetical effects. Most low-energy
tests of charge symmetry showed that the symmetry holds within about
1\% in the reaction amplitudes~\cite{miller90}.

At present, charge symmetry is assumed to be valid in almost all
phenomenological parton distributions, because there is no direct
experimental confirmation pointing to a substantial CSV in the
parton distributions. As we know, CSV effect is small and can be
hardly separated clearly from the strangeness asymmetry, besides
that CSV effect also often mixes with the flavor symmetry
violation~(FSV) effect, i.e., the asymmetry between $\bar{u}$ and
$\bar{d}$ quarks in the nucleon~\cite{ma92}. Until recently, with
the development of high-energy deep-inelastic scattering
experiments, the detailed information regarding the structure of the
nucleon is known much better. The famous experiment of asymmetry for
$\bar{u}$ and $\bar{d}$ distributions in the nucleon sea, carried
out first by the NMC group~\cite{a97,nmc}, enabled a better
determination of the Gottfried Sum~\cite{gsr}, which was predicted
to be 1/3 with the assumptions of charge symmetry and flavor
symmetry. Later the flavor asymmetric sea was also confirmed by the
pp and pD Drell-Yan processes~\cite{drellyan} and by the
semi-inclusive electroproduction at HERMS~\cite{herms}. All these
experimental measurements have been interpreted as evidences of FSV,
but they can be also explained by a large CSV effect if the FSV
effect is neglected~\cite{ma92}. It is remarkable that the possible
violation of charge symmetry~(CS) has attracted attention again,
because it might be closely related with the NuTeV anomaly. The
earliest estimation of CSV effect relating to the Weinberg angle is
made by Sather~\cite{sather92}. He first pointed out that the charge
symmetry violation~(CSV) must be understood when a high-precision
value of $\sin^{2}\theta_{w}$ is extracted from deep inelastic
neutrino scattering, and gave the correction to $\sin^{2}\theta_{w}$
around 0.002 within the nonperturbative framework of quark model.
Qualitatively, a similar conclusion was obtained within the bag
model~\cite{rodionov94} by including a number of effects neglected
in Ref.~\cite{sather92}. Both of these models predict that the
``majority" quark distributions satisfy charge symmetry violation
within about 1\%, while the ``minority" quark distributions are
predicted to violate CS around 5\% or more at large $x$. In
Ref.~\cite{benesh98}, the authors combined the approaches of
Ref.~\cite{sather92} and Ref.~\cite{rodionov94} to examine the
violation of CS in the valence and sea quark distributions of the
nucleon, and found that the size of CSV effect is large in the
valence quark distributions~(same as the conclusion of
Refs.~\cite{sather92,rodionov94,benesh97}) and too small in the
nucleon sea to have significant contribution to any observable.
Later, Davidson and Burkardt~\cite{davidson97} employed the
convolution approach~\cite{dunne86} to estimate the charge symmetry
breaking effects and gave the size of the correction to the Weinberg
angle. Boros~\cite{boros98} and his collaborators presented a
serious challenge to CS by comparing the structure functions
$F_{2}^{\nu}(x, Q^{2})$ from neutrino-induced charged-changing
reactions by the CCFR Collaboration~\cite{sbr97} and the structure
functions $F_{2}^{\mu}(x, Q^{2})$ from charged lepton DIS by the NMC
collaboration~\cite{nmc}, and placed the upper limits in the
magnitude of CSV. After that, there has been a series of discussions
about CSV contribution to the NuTeV discrepancy. Londergan and
Thomas~\cite{lt03} investigated the CSV effect, and suggested that
it is largely independent of parton distribution functions (PDFs)
and should reduce roughly 30\% of the discrepancy between the NuTeV
measurement and the world accepted value of $\sin^{2}\theta_{w}$.
Also, the MRST~\cite{mrst} group obtained a phenomenological
evaluation of PDFs including isospin violating by widely fitting a
variety of high energy experimental data.
%If using the MRST value of sea- and valence-quark
%distributions, one can gained the CSV effects which would reduce
%about $\frac{1}{3}$ of the NuTeV deviations.
On the contrary, Cao and Signal~\cite{cs00} calculated the
nonperturbative effect of CSV within the framework of meson cloud
model and showed no contribution to the NuTeV anomaly. And recently,
the contribution to the valence isospin violation stemming from
dynamical~(radiative) QED effect was investigated~\cite{martin05}
and the size of CSV effect ~\cite{gluck05} is similar to those
calculated within the bag model.  In this paper, we
%have an insight into the origin and
analyze qualitatively the CSV effect within the light-cone
meson-baryon fluctuation model~\cite{bm96}, and show that the
contribution is too small to affect the measurement of the Weinberg
angle in the neutrino scattering, unless a larger difference between
the internal momentum scales, $\alpha_{p}$ of the proton and
$\alpha_{n}$ of the neutron, is taken into account.

\section{The correction of CSV effect to the NuTeV anomaly}

In the realistic reaction, the Paschos-Wolfenstein
relation~\cite{pash73} must be corrected by the possible effects of
non-isoscalar target, $s(x)$-$\bar{s}(x)$ and $c(x)$-$\bar{c}(x)$
asymmetries, and charge symmetry violation in the nucleon sea. In
this section, we will give a revised expression for the P-W relation
with CSV effects. The procedure is similar to that for the
asymmetric $s$-$\bar{s}$ momentum distributions in the nucleon sea
in Ref.~\cite{dm04,dxm05}, about which we make a brief review here.
As we know, $\sigma^{\nu N}_{NC}$ ($\sigma^{\bar{\nu} N}_{NC}$) in
Eq.~(\ref{ratio}) is the integral of differential cross section over
$x$ and $y$ for neutral-current reactions induced by
neutrino~(antineutrino) on nucleon target, and it is the same for
$\sigma^{\nu N}_{CC}$~($\sigma^{\bar{\nu} N}_{CC}$). The most
general form of the differential cross section for neutral-current
interactions initialed by (anti-)neutrino is~\cite{lt98}:
\begin{eqnarray}
\frac{\textmd{d}^{2}\sigma^{\nu(\bar{\nu})}_{NC}}{\textmd{d}
x\textmd{d} y}&=&\pi
s\left(\frac{\alpha}{2\sin^{2}\theta_{w}\cos^{2}\theta_{w}M^{2}_{Z}}
\right)^{2}\left(\frac{M^{2}_{Z}}{M^{2}_{Z}+Q^{2}}\right)^{2} \bigg{[}xyF^{Z}_{1}(x,Q^{2}) \nonumber\\
&&
+\left(1-y-\frac{xym^{2}_{N}}{s}\right)F^{Z}_{2}(x,Q^{2})\pm\left(y-\frac{y^{2}}{2}\right)xF^{Z}_{3}(x,Q^{2})\bigg{]}.
\end{eqnarray}
Similarly, we can have the cross section for (anti-)neutrino-nucleon
charged-current reaction~\cite{lt98},
%the form~\cite{lt98} is:
\begin{eqnarray}
\frac{\textmd{d}^{2}\sigma^{\nu(\bar{\nu})}_{CC}}{\textmd{d} x
\textmd{d} y}&=&{\pi}s\left(\frac{\alpha}{2\sin^{2}\theta_{w}
M^{2}_{W}}\right)^{2}\left(\frac{M^{2}_{W}}{M^{2}_{W}+Q^{2}}\right)^{2}
 \bigg{[}xyF^{W^{\pm}}_{1}(x,Q^{2}) \nonumber\\
&& +\left(1-y-\frac{xym^{2}_{N}}{s}\right)
F^{W^{\pm}}_{2}(x,Q^{2})\pm\left(y-\frac{y^{2}}{2}\right)xF^{W^{\pm}}_{3}(x,Q^{2})
\bigg{]},
\end{eqnarray}
where $M_{Z}$ and $M_{W}$ are the masses of the neutral-  and
charged-current interacting weak vector bosons, respectively,
$\theta_{w}$ is the Weinberg angle, $x=Q^{2}/2p\cdot q$, $y=p\cdot
q/p\cdot k$, $Q^{2}=-q^{2}$ is the square of the four momentum
transfer for the reaction, $k$ ($p$) is the momentum of the initial
state for neutrino or antineutrino (nucleon), and $s=(k+p)^{2}$.
Besides these, $F^{Z(W^{\pm}) p}_{i}(x,Q^{2})$ are the structure
functions on the proton ($p$), which only depend on $x$ as
$Q^{2}\rightarrow\infty$, and are written in terms of the parton
distributions as~\cite{lt98}
\begin{eqnarray}
\lim_{Q^{2}\rightarrow\infty}F^{Zp}_{1}(x,Q^{2})&=&\frac{1}{2}\bigg{[}\bigg{(}(f^{u}_{V})^{2}+(f^{u}_{A})^{2}\bigg{)}
\bigg{(}u^{p}(x)+\bar{u}^{p}(x)
    +c^{p}(x)+\bar{c}^{p}(x)\bigg{)} \nonumber\\&&+\bigg{(}(f^{d}_{V})^{2}+(f^{d}_{A})^{2}\bigg{)}
    \bigg{(}d^{p}(x)+\bar{d}^{p}(x)
    +s^{p}(x)+\bar{s}^{p}(x)\bigg{)}\bigg{]},\nonumber\\
\lim_{Q^{2}\rightarrow\infty}F^{Zp}_{3}(x,Q^{2})&=&2\bigg{[}f^{u}_{V}f^{u}_{A}\bigg{(}u^{p}(x)-
    \bar{u}^{p}(x)+c^{p}(x)-\bar{c}^{p}(x)\bigg{)}\nonumber\\&&+f^{d}_{V}f^{d}_{A}\bigg{(}d^{p}(x)-
    \bar{d}^{p}(x)+s^{p}(x)-\bar{s}^{p}(x)\bigg{)}\bigg{]}. \nonumber\\
F^{Zp}_{2}(x,Q^{2})&=&2xF^{Zp}_{1}(x,Q^{2}).\label{struc}
\end{eqnarray}
The structure functions of charged-current in above equations have
the forms:
\begin{eqnarray}
\lim_{Q^{2}\rightarrow\infty}F^{W^{+}p}_{1}(x,Q^{2})&=&d^{p}(x)+\bar{u}^{p}(x)+s^{p}(x)+\bar{c}^{p}(x),\nonumber\\
\lim_{Q^{2}\rightarrow\infty}F^{W^{-}p}_{1}(x,Q^{2})&=&u^{p}(x)+\bar{d}^{p}(x)+\bar{s}^{p}(x)+c^{p}(x),\nonumber\\
\frac{1}{2}\lim_{Q^{2}\rightarrow\infty}F^{W^{+}p}_{3}(x,Q^{2})&=&d^{p}(x)-\bar{u}^{p}(x)+s^{p}(x)-\bar{c}^{p}(x),\nonumber\\
\frac{1}{2}\lim_{Q^{2}\rightarrow\infty}F^{W^{-}p}_{3}(x,Q^{2})&=&u^{p}(x)-\bar{d}^{p}(x)-\bar{s}^{p}(x)+c^{p}(x),\nonumber\\
F^{W^{\pm}p}_{2}(x,Q^{2})&=&2xF^{W^{\pm}p}_{1}(x,Q^{2}).\label{stru}
\end{eqnarray}
One can obtain the structure functions for the neutron~$n$ by
replacing the superscripts $p\rightarrow$$n$ everywhere in
Eqs.~(\ref{struc}) and~(\ref{stru}). In Eq.~(\ref{struc}),
$f^{u}_{V}$, $f^{u}_{A}$, $f^{d}_{V}$ and $f^{d}_{A}$ are vector and
axial-vector couplings:
$$f^{u}_{V}=\frac{1}{2}-\frac{4}{3}\sin^{2}\theta_{w},  \ \ f^{u}_{A}=\frac{1}{2},$$
$$f^{d}_{V}=-\frac{1}{2}+\frac{2}{3}\sin^{2}\theta_{w}, \ \ f^{d}_{A}=-\frac{1}{2}.$$ Charge
symmetry means that:
\begin{eqnarray}
   d^{n}(x)&=&u^{p}(x),\nonumber\\
   u^{n}(x)&=&d^{p}(x),\nonumber\\
  s^{n}(x)&=&s^{p}(x)=s(x),\nonumber\\
  c^{n}(x)&=&c^{p}(x)=c(x), \label{pdis}
\end{eqnarray}
and $c(x)=\bar{c}(x)$,$s(x)=\bar{s}(x)$. Thus, with charge
symmetry violation and assumption of $s(x)=\bar{s}(x)$, the
structure functions for isoscalar target in the neutrino
charged-current reaction are given by
\begin{eqnarray}
\lim_{Q^{2}\rightarrow\infty}F^{W^{+}N}_{1}(x,Q^{2})&=&\frac{1}{2}
\bigg{[}d^{p}(x)+\bar{d}^{p}(x)+u^{p}(x)+\bar{u}^{p}(x)+2s(x)+2\bar{c}(x)-\delta u(x)-\delta\bar{d}(x)\bigg{]},\nonumber\\
\lim_{Q^{2}\rightarrow\infty}F^{W^{-}N}_{1}(x,Q^{2})&=&\frac{1}{2}
\bigg{[}d^{p}(x)+\bar{d}^{p}(x)+u^{p}(x)+\bar{u}^{p}(x)+2\bar{s}(x)+2c(x)-\delta d(x)-\delta\bar{u}(x)\bigg{]},\nonumber\\
\lim_{Q^{2}\rightarrow\infty}F^{W^{+}N}_{3}(x,Q^{2})&=&
d^{p}(x)+u^{p}(x)-\bar{u}^{p}(x)-\bar{d}^{p}(x)+2s(x)-2\bar{c}(x)-\delta u(x)+\delta\bar{d}(x),\nonumber\\
\lim_{Q^{2}\rightarrow\infty}F^{W^{-}N}_{3}(x,Q^{2})&=&
d^{p}(x)+u^{p}(x)-\bar{u}^{p}(x)-\bar{d}^{p}(x)-2\bar{s}(x)+2c(x)-\delta d(x)+\delta\bar{u}(x),\nonumber\\
F^{W^{\pm}N}_{2}(x,Q^{2})&=&2xF^{W^{\pm}N}_{1}(x,Q^{2}),\label{str}
\end{eqnarray}
where
$F^{W^{+}N}_{i}(x,Q^{2})=\frac{1}{2}\bigg{(}F^{W^{+}p}_{i}(x,Q^{2})+F^{W^{+}n}_{i}(x,Q^{2})\bigg{)}$.
And the forms of structure functions for the neutral-current also
can be obtained in the same way:
\begin{eqnarray}
\lim_{Q^{2}\rightarrow\infty}F^{ZN}_{1}(x,Q^{2})&=&\frac{1}{2}
\bigg{[}\bigg{(}(f^{u}_{V})^{2}+(f^{u}_{A})^{2}\bigg{)}\bigg{(}d^{p}(x)+\bar{d}^{p}(x)+u^{p}(x)+\bar{u}^{p}(x)
+2c(x)+2\bar{c}(x)\nonumber\\&-&\delta
d(x)-\delta\bar{d}(x)\bigg{)}+\bigg{(}(f^{d}_{V})^{2}+(f^{d}_{A})^{2}\bigg{)}\bigg{(}d^{p}(x)+\bar{d}^{p}(x)+u^{p}(x)\nonumber\\&+&\bar{u}^{p}(x)
+2s(x)+2\bar{s}(x)-\delta u(x)-\delta\bar{u}(x)\bigg{)}\bigg{]},\nonumber\\
 \lim_{Q^{2}\rightarrow\infty}F^{ZN}_{3}(x,Q^{2})&=&f^{u}_{V}f^{u}_{A}
\bigg{[}d^{p}_{v}(x)+u^{p}_{v}(x)-\delta
d_{v}(x)+2c_{v}(x)\bigg{]},\nonumber\\&+&f^{d}_{V}f^{d}_{A}
\bigg{[}d^{p}_{v}(x)+u^{p}_{v}(x)-\delta u_{v}(x)+2s_{v}(x)\bigg{]},\nonumber\\
F^{ZN}_{2}(x,Q^{2})&=&2xF^{ZN}_{1}(x,Q^{2}),\label{str}
\end{eqnarray}
where $q^{p}_{v}=q^{p}(x)-\bar{q}^{p}(x)$ is the valence
distribution of flavor $q$ quarks in the proton $p$.  Thus, using
the structure functions above, one can derive the modified P-W
relation with the CSV effects:
\begin{eqnarray}
  R^{-}_{N}=\frac{\sigma^{\nu N}_{NC}-\sigma^{\bar{\nu}N}_{NC}}{\sigma^{\nu N}_{CC}-\sigma^{\bar{\nu}N}_{CC}}
= R^{-}-\delta R^{-}_{csv},\label{correction}
\end{eqnarray}
where $\delta R^{-}_{csv}$ is the correction brought by the effects
of CSV to the naive P-W relation $R^{-}$ and has the following form:
\begin{eqnarray}
  \delta
  R^{-}_{csv}=\bigg{(}\frac{1}{2}-\frac{7}{6}\sin^{2}\theta_{w}\bigg{)}\frac{\int^{1}_{0}x\bigg{[}\delta d_{v}(x)-\delta u_{v}(x)\bigg{]}\d x}
  {\int^{1}_{0}x\bigg{[}u_{v}(x)+d_{v}(x)\bigg{]}\d x},\label{rs}
\end{eqnarray}
with
\begin{eqnarray}
\delta u_{v}(x)&=&\delta u(x)-\delta\bar{u}(x),\nonumber\\
\delta d_{v}(x)&=&\delta d(x)-\delta\bar{d}(x),\nonumber\\
\delta u(x)&=&u^{p}(x)-d^{n}(x),\nonumber\\
\delta \bar{u}(x)&=&\bar{u}^{p}(x)-\bar{d}^{n}(x).
\end{eqnarray}
From this equation, we find that the violation of CS should bring
correction to $\sin^{2}\theta_{w}$, at least as shown in the
formalism. In the remaining section, we will give a detailed
calculation about $\delta R^{-}_{csv}$  by using the light-cone
meson-baryon fluctuation model and the light-cone
quark-spectator-diquark model.

\section{Charge symmetry violation
\label{Sec:Three}}

In this section, we will perform the calculation of CSV by adopting
the mechanisms of the light-cone meson-baryon fluctuation
model~\cite{bm96} and the light-cone quark-spectator-diquark
model~\cite{ma96}. In the light-cone meson-baryon fluctuation model,
the hadronic wave function can be expressed by a series of
light-cone wave functions multiplied by the Fock states. Usually,
the proton wave function can be written as:
\begin{eqnarray}
   \left|p\right\rangle=\left|uud\right\rangle\Psi_{uud/p}+ \left|uudg\right\rangle\Psi_{uudg/p}+\sum_{q\bar{q}}
   \left|uudq \bar{q}\right\rangle\Psi_{uudq\bar{q}/p}+\cdots.
\end{eqnarray}
Here, we adopt the approximation in Ref.~\cite{bm96},
%made by Brodsky and Ma~\cite{bm96},
in which the intrinsic sea quarks of the proton are estimated by the
meson-baryon fluctuations:
\begin{eqnarray}
&&p(uud)\rightarrow\pi^{+}(u\bar{d})n(udd),\nonumber\\
&&p(uud)\rightarrow\pi^{+}(u\bar{d})\Delta^{0}(udd),\nonumber\\
&&p(uud)\rightarrow p(uud)\pi^{0}\bigg{(}\frac{1}{\sqrt{2}}[u\bar{u}-d\bar{d}]\bigg{)},\nonumber\\
&&\cdots\cdots
\end{eqnarray}
and estimate the violation of charge symmetry coming from these
fluctuations. In the same way, the neutron sea can be obtained, so
that we can evaluate the relative probabilities of two different
meson-baryon fluctuation states by comparing their relevant
off-shell light-cone energies. In this paper, we choose the
light-cone Gaussian type wave function as in Ref.~\cite{ma97} as a
two-body wave function:
\begin{eqnarray}
   \Psi(M^{2})=A_{G}\exp\bigg{[}-(M^{2}-m^{2}_{N})/8\alpha^{2}_{D}\bigg{]},\label{si}
\end{eqnarray}
where $M^{2}=\sum^{2}_{i=1}\frac{\mathbf{k}_{\perp
i}^{2}+m_{i}^{2}}{x_{i}}$ is the invariant mass square for the
meson-baryon state, $m_{N}$ is the physical mass of the nucleon,
$\alpha$ is the characteristic internal momentum scale,
$\mathbf{k}_{\bot}$ is the internal transversal momentum, and
$A_{G}$ is the normalization constant.

In this paper, we recalculate the relative probabilities of
$p\rightarrow\pi^{+}n$ to $n\rightarrow\pi^{-}p$ and find that the
ratios:
$r^{\pi}_{p/n}=P(p\rightarrow\pi^{+}n)/P(n\rightarrow\pi^{-}p)$, are
very different for the different inputs of $\alpha_{p}$ and
$\alpha_{n}$ and there is an excess of $n\rightarrow\pi^{-}p$ over
$p\rightarrow\pi^{+}n$ fluctuation.
%with assuming $P(p\rightarrow\pi^{+}n)\simeq P( n\rightarrow\pi^{-}p)$.
For example: there is an excess of 0.2\% of $n\rightarrow\pi^{-}p$
over $p\rightarrow\pi^{+}n$ fluctuation with assuming
$P(p\rightarrow\pi^{+}n)\simeq P( n\rightarrow\pi^{-}p)\simeq0.15$,
when $\alpha=330$~MeV for both proton and  neutron,
$m_{p}=938.27$~MeV and $m_{n}=939.57$~MeV as the physical masses of
proton and neutron, respectively. We also reexamine the case
calculated in Ref.~\cite{ma97}: the ratio of
$r^{\pi}_{p/n}=P(p\rightarrow\pi^{+}n)/P(n\rightarrow\pi^{-}p)=0.820$
which means that there is an excess of 3\% of $n\rightarrow\pi^{-}p$
over $p\rightarrow\pi^{+}n$ fluctuation, when $\alpha=200$~MeV for
the proton and $\alpha=205$~MeV for the neutron~(considering that
the Coulomb attraction between $\pi^{-}$ and $p$ in the fluctuation
$n\rightarrow\pi^{-}p$ may require larger relative motions of pions
than in the fluctuation state $p\rightarrow\pi^{+}n$). The
calculations of other cases are showed in Table~\ref{csv}. Besides
that we also make an estimation about the ratios for probabilities
between other fluctuations to $p(n)\rightarrow\pi^{+(-)}n(p)$
fluctuation and find that the ratios are relative smaller than the
ratio $p\rightarrow\pi^{+}n$ to $n\rightarrow\pi^{-}p$. In this
work, we neglect the effects from other fluctuations, mainly
because, firstly, the relative ratios of other fluctuations to the
fluctuation $p(n)\rightarrow\pi^{+(-)}n(p)$ is small; secondly, the
neutral fluctuations to chargeless meson $\pi^{0}$ $et.~al$ hardly
contribute to the CSV effect; thirdly, there is only about 0.1\%
excess of $p\rightarrow K^{+}\Lambda$ over $n\rightarrow
K^{0}\Lambda$ fluctuation, which is also rather small compared with
the fluctuation $p\rightarrow\pi^{+}n$ to $n\rightarrow\pi^{-}p$,
with $P(p\rightarrow K^{+}\Lambda)\simeq P(n\rightarrow
K^{0}\Lambda)\simeq0.05$. Thus, we can obtain a crude model
estimation for the excess of $n\rightarrow\pi^{-}p$ over
$p\rightarrow\pi^{+}n$ fluctuation states, which mainly arises from
the small difference $m_{n}-m_{p}=1.3$~MeV. This excess seems to be
an important source for CSV in the valence and sea quark
distributions between the proton and the neutron within the
light-cone meson-baryon fluctuation model. When we only consider the
fluctuation $n\rightarrow\pi^{-}p$ and $p\rightarrow\pi^{+}n$, the
CSV in the valence and sea quark distributions can be obtained by
the $u$ and $d$ distributions in the proton and the neutron:
\begin{eqnarray}
u^{p}(x)&=&\int^{1}_{x}\frac{\d
y}{y}\bigg{[}f_{\pi^{+}/\pi^{+}n}(y)f_{u/u\bar{d}}\bigg{(}\frac{x}{y}\bigg{)}+f_{n/\pi^{+}n}(y)f_{u/udd}\bigg{(}\frac{x}{y}\bigg{)}\bigg{]},\nonumber\\
d^{n}(x)&=&\int^{1}_{x}\frac{\d
y}{y}\bigg{[}f_{\pi^{-}/\pi^{-}p}(y)f_{d/d\bar{u}}\bigg{(}\frac{x}{y}\bigg{)}+f_{p/\pi^{-}p}(y)f_{d/uu
d}\bigg{(}\frac{x}{y}\bigg{)}\bigg{]}.
\end{eqnarray}
where
\begin{eqnarray}
f_{\pi^{+}/\pi^{+}n}\left(\frac{x}{y}\right)&=&\int^{+\infty}_{-\infty}\d
\mathbf{k}_{\bot}\bigg{|}A_{p}\exp\bigg{[}-\frac{1}{8\alpha^{2}}
\bigg{(}\frac{m_{\pi}^{2}+\mathbf{k}_{\bot}^{2}}{x}+\frac{m_{n}^{2}+\mathbf{k}_{\bot}^{2}}{1-x}-m_{n}^{2}\bigg{)}\bigg{]}\bigg{|}^{2},\nonumber\\
f_{\pi^{-}/\pi^{-}p}\left(\frac{x}{y}\right)&=&\int^{+\infty}_{-\infty}\d
\mathbf{k}_{\bot}\bigg{|}A_{n}\exp\bigg{[}-\frac{1}{8\alpha^{2}}
\bigg{(}\frac{m_{\pi}^{2}+\mathbf{k}_{\bot}^{2}}{x}+\frac{m_{p}^{2}+\mathbf{k}_{\bot}^{2}}{1-x}-m_{p}^{2}\bigg{)}\bigg{]}\bigg{|}^{2},\nonumber\\
f_{u/u\bar{d}}\left(\frac{x}{y}\right)&=&f_{d/d\bar{u}}\left(\frac{x}{y}\right)=\int^{+\infty}_{-\infty}\d
\mathbf{k}_{\bot}\bigg{|}A_{u}\exp\bigg{[}-\frac{1}{8\alpha^{2}}
\bigg{(}\frac{m_{u}^{2}+\mathbf{k}_{\bot}^{2}}{x}+\frac{m_{d}^{2}+\mathbf{k}_{\bot}^{2}}{1-x}\bigg{)}\bigg{]}\bigg{|}^{2},\nonumber\\
f_{u/uud}\left(\frac{x}{y}\right)&=&f_{d/udd}\left(\frac{x}{y}\right)=\int^{+\infty}_{-\infty}\d
\mathbf{k}_{\bot}\bigg{|}A_{D}\exp\bigg{[}-\frac{1}{8\alpha^{2}}
\bigg{(}\frac{m_{u(d)}^{2}+\mathbf{k}_{\bot}^{2}}{x}+\frac{m_{D}^{2}+\mathbf{k}_{\bot}^{2}}{1-x}\bigg{)}\bigg{]}\bigg{|}^{2}.~~~~~
\end{eqnarray}
In the same way, we can also obtain the other distributions
$u^{n}(x)$, $d^{p}(x)$, $\bar{u}^{p(n)}(x)$ and $\bar{d}^{p(n)}(x)$,
from which one can obtain the result of CSV. As for the $u$ and $d$
valence quark distributions, one can calculate them in the
light-cone quark-spectator-diquark model~\cite{ma96}:
\begin{eqnarray}
   u_{v}(x)&=&\frac{1}{2}a_{S}(x)+\frac{1}{6}a_{V}(x),\nonumber\\
   d_{v}(x)&=&\frac{1}{3}a_{V}(x),
\end{eqnarray}
where $a_{D}(x)$ ($D=S$ or $V$, with $S$ standing for scalar
diquark Fock state and $V$ standing for vector diquark state)
denotes that the amplitude for the quark $q$ is scattered while
the spectator is in diquark state $D$ \cite{kbb}, and can be
written as:
\begin{equation}
   a_{D}(x)\propto\int[\d\mathbf{k}_{\bot}]\bigg{|}\Psi_{D}(x,\mathbf{k}_{\bot})\bigg{|}^{2}.
\end{equation}
Here, we adopt the light-cone momentum wave function of the
quark-spectator-diquark model
\begin{equation}
\Psi_{D}(x,\mathbf{k}_{\bot})=A_{D}\exp\bigg{[}-\frac{1}{8\alpha^{2}_{D}}\bigg{(}\frac{m^{2}_{q}+\mathbf{k}_{\bot}^{2}}
{x}+\frac{m^{2}_{D}+\mathbf{k}_{\bot}^{2}}{1-x}\bigg{)}\bigg{]}.
\end{equation}
Using above equations, one can have the CSV effect and the
correction to the NuTeV anomaly, which are given in Table~\ref{csv}
with the parameters: $m^{\pm}_{\pi}=139.57$~MeV, $m_{S}=600$~MeV,
$m_{V}=900$~MeV, $m_{p}=938.27$~MeV and $m_{n}=939.57$~MeV.
\begin{table}[!htbp]\caption {\label{csv}The correction to the NuTeV anomaly for different parameters}
\begin{center}
\begin{tabular}{|c|c|c|c|}
  \hline
  \hline
  mass~(MeV) & $\alpha$~(MeV) & $r_{p/n}$ & $ \delta R^{-}_{csv}$  \\
  \hline
  $m_{u(\bar{u})}=m_{d(\bar{d})}=330$ & $\alpha_{p}=\alpha_{n}=330$ & 0.986 & $-0.45\times10^{-5}$ \\
  $m_{u(\bar{u})}=m_{d(\bar{d})}=330$ & $\alpha_{p}=330$, $\alpha_{n}=335$ & 0.924 & $-0.5\times10^{-4}$\\
  $m_{u(\bar{u})}=m_{d(\bar{d})}=330$ & $\alpha_{p}=\alpha_{n}=220$ & 0.971 & $-0.38\times10^{-5}$\\
  $m_{u(\bar{u})}=m_{d(\bar{d})}=330$ & $\alpha_{p}=200$, $\alpha_{n}=205$ & 0.820 &  $-0.6\times10^{-4}$ \\
  $m_{u(\bar{u})}=m_{d(\bar{d})}=330$ & $\alpha_{p}=200$, $\alpha_{n}=210$  & 0.701 & $-0.12\times10^{-3}$ \\
  $m_{u(\bar{u})}=330$, $m_{d(\bar{d})}= 334$ & $\alpha_{p}=\alpha_{n}=330$ & 0.986 & $-0.37\times10^{-4}$ \\
  $m_{u(\bar{u})}=330$, $m_{d(\bar{d})}= 334$ & $\alpha_{p}=330$, $\alpha_{n}=335$  & 0.924 & $-0.84\times10^{-4}$\\
  $m_{u(\bar{u})}=330$, $m_{d(\bar{d})}= 334$ & $\alpha_{p}=\alpha_{n}=220$ & 0.971 & $-0.42\times10^{-4}$\\
  $m_{u(\bar{u})}=330$, $m_{d(\bar{d})}= 334$ & $\alpha_{p}=200$, $\alpha_{n}=205$ & 0.820 & $-0.83\times10^{-4}$ \\
  $m_{u(\bar{u})}=330$, $m_{d(\bar{d})}= 334$ & $\alpha_{p}=200$, $\alpha_{n}=210$ & 0.701 & $-0.15\times10^{-3}$\\
    \hline
\end{tabular}
\end{center}
\end{table}
In the Table~\ref{csv}, $\alpha_{p(n)}$ denotes the internal
momentum scale for $p\rightarrow\pi^{+}n(n\rightarrow\pi^{-}p)$.
From the Table, we find that the value of $\alpha$ has an important
impact on the ratio $r_{p/n}$ and $\delta R^{-}_{csv}$, while the
role of mass difference is not so important. When the difference of
$\alpha_{p}$ and $\alpha_{n}$ is larger, the ratio is larger and
$\delta R^{-}_{csv}$ has larger contribution to the measurement of
Weinberg angle. However, as pointed by Miller in
Ref.~\cite{miller98}, the effect of CSV on the nucleon might be
overestimated quantitatively in case of a larger difference between
$\alpha_{p}$ and $\alpha_{n}$, though such a difference is
physically reasonable.

\section{conclusion}

As we have known, charge symmetry (CS) is an extremely well
respected symmetry, and there are only some upper limit estimates on
charge symmetry violation (CSV) of the parton distributions at
present. Although there have been so many theoretical discussions
about it, unfortunately, there is no direct way to verify the
theoretical predictions. In this work, we calculated the CSV effect
by using the light-cone meson-baryon fluctuation model and estimated
the contribution from the CSV effect to the NuTeV anomaly. From
Table~\ref{csv}, we found that the correction to the NuTeV
measurement is very small in magnitude unless the larger difference
between internal momentum scales, i.e., $\alpha_{p}$ of the proton
and $\alpha_{n}$ of the neutron, is considered. Therefore it might
be necessary to study this effect more carefully, so that more
reliable prediction can be made. As for whether the CSV effect is
sizable or not, it would need further studies both theoretically and
experimentally. Therefore more precision experiments should be
carried out in the future to provide more detailed information on
the parton structure of the nucleon.

\section{Acknowledgments}

This work is partially supported by National Natural Science
Foundation of China (Nos.~10505001, 10421503, 10575003, 10528510),
by the Key Grant Project of Chinese Ministry of Education
(No.~305001), and by the Research Fund for the Doctoral Program of
Higher Education (China).

\end{document}